\pdfoutput=1
\documentclass[iop]{emulateapj}

\usepackage{epsfig}
\usepackage{longtable}
\usepackage[breaklinks]{hyperref}
\usepackage{multirow}
\usepackage{mypack}
\def\bfref{}

\begin{document}

\title{Timing of Five PALFA-Discovered Millisecond Pulsars}

\author{K.~Stovall$^{1}$, B.~Allen$^{2,3,4}$, S.~Bogdanov$^{5}$,
A.~Brazier$^{6,7}$, F.~Camilo$^{8}$, F.~Cardoso$^{9}$, S.~Chatterjee$^{6}$,
J.~M.~Cordes$^{6}$, F.~Crawford$^{10}$, J.~S.~Deneva$^{11}$, R.~Ferdman$^{12}$,
P.~C.~C.~Freire$^{14}$, J.~W.~T.~Hessels$^{15,16}$,
F.~Jenet$^{17}$, D.~L.~Kaplan$^{5}$, C.~Karako-Argaman$^{12}$, V.~M.~Kaspi$^{12}$, B.~Knispel$^{2,3}$,
R.~Kotulla$^{4,18}$, P.~Lazarus$^{14}$, K.~J.~Lee$^{19}$, J.~van~Leeuwen$^{15,16}$,
R.~Lynch$^{20}$, A.~G.~Lyne$^{21}$, E.~Madsen$^{12}$, M.~A.~McLaughlin$^{9}$,
C.~Patel$^{12}$, S.~M.~Ransom$^{20}$, P.~Scholz$^{12}$, X.~Siemens$^{5}$,
I.~H.~Stairs$^{22,13}$, B.~W.~Stappers$^{21}$, J.~Swiggum$^{5}$, W.~W.~Zhu$^{14}$, A.~Venkataraman$^{23}$}


\footnotetext[1]{Department of Physics and Astronomy, University of New Mexico, Albuquerque, NM, USA; stovall.kevin@gmail.com}
\footnotetext[2]{Max-Planck-Institut f\"{u}r  Gravitationsphysik,  D-30167  Hannover, Germany}
\footnotetext[3]{Leibniz Universit\"{a}t Hannover, D-30167 Hannover, Germany}
\footnotetext[4]{Physics Dept., Univ. of Wisconsin - Milwaukee, Milwaukee WI 53211, USA}
\footnotetext[5]{Columbia Astrophysics Laboratory, Columbia Univ., New York, NY 10027, USA}
\footnotetext[6]{Dept. of Astronomy, Cornell Univ., Ithaca, NY 14853, USA}
\footnotetext[7]{Center for Advanced Computing, Cornell Univ., Ithaca, NY 14853, USA}
\footnotetext[8]{SKA South Africa, Pinelands, 7405, South Africa}
\footnotetext[9]{Dept. of Physics, West Virginia Univ., Morgantown, WV 26506, USA}
\footnotetext[10]{Dept. of Physics and Astronomy, Franklin and Marshall College, Lancaster, PA 17604-3003, USA}
\footnotetext[11]{National Research Council, resident at the Naval Research Laboratory, 4555 Overlook Ave. SW, Washington, DC 20375, USA}
\footnotetext[12]{Dept. of Physics, McGill Univ., Montreal, QC H3A 2T8, Canada}
\footnotetext[13]{McGill Space Institute, McGill Univ., Montreal, QC H3A 2T8, Canada}
\footnotetext[14]{Max-Planck-Institut f\"{u}r Radioastronomie, Auf dem H\"{u}gel 69, 53131 Bonn, Germany}
\footnotetext[15]{ASTRON, the Netherlands Institute for Radio Astronomy, Postbus 2, 7990 AA, Dwingeloo, The Netherlands}
\footnotetext[16]{Anton Pannekoek Institute for Astronomy, Univ. of Amsterdam, Science Park 904, 1098, XH Amsterdam, The Netherlands}
\footnotetext[17]{Center for Advanced Radio Astronomy, Univ. of Texas at Brownsville, TX 78520, USA}
\footnotetext[18]{Department of Astronomy, University of Wisconsin-Madison, 475 N Charter St, Madison, WI, 53707, USA}
\footnotetext[19]{Kavli Institute for Astronomy and Astrophysics, Peking Univ., Beijing 100871, P.R. China}
\footnotetext[20]{NRAO, Charlottesville, VA 22903, USA}
\footnotetext[21]{Jodrell Bank Centre for Astrophysics, Univ. of Manchester, Manchester, M13 9PL, UK}
\footnotetext[22]{Dept. of Physics and Astronomy, Univ. of British Columbia, Vancouver, BC V6T 1Z1, Canada}
\footnotetext[23]{Arecibo Observatory, HC3 Box 53995, Arecibo, PR 00612}

\begin{abstract}
We report the discovery and timing results for five millisecond
pulsars (MSPs) from the Arecibo PALFA survey: PSRs J1906+0055,
J1914+0659, J1933+1726, J1938+2516, and J1957+2516. Timing
observations of the 5 pulsars were conducted with the Arecibo and
Lovell telescopes for time spans ranging from 1.5 to 3.3 yr. All of
the MSPs except one (PSR J1914+0659) are in binary systems with low
eccentricities. PSR J1957+2516 is likely a redback pulsar, {\bfref with a
$\sim 0.1~M_{\odot}$ companion and possible eclipses that last 
$\sim 10$\% of the orbit. The position of PSR J1957+2516 is also
coincident with a NIR source.}
All 5 MSPs are distant ($> 3.1$ kpc) as determined from their
dispersion measures, and none of them show evidence of $\gamma$-ray
pulsations in a search of {\it Fermi Gamma-Ray Space Telescope} data.
These 5 MSPs bring the total number of MSPs discovered
by the PALFA survey to 26 and further demonstrate the power of this
survey in finding distant, highly dispersed MSPs deep in the Galactic plane.
\end{abstract}
\maketitle

\section{Introduction}\label{sec:intro}
In recent years, several large-scale pulsar surveys have been undertaken to
search for new pulsars~\citep{2006ApJ...637..446C,2010MNRAS.409..619K,
2013ApJ...763...80B,2013ApJ...775...51D,2013MNRAS.435.2234B,
2014ApJ...791...67S,2014A&A...570A..60C}.
One of the drivers for such surveys is the discovery of millisecond
pulsars (MSPs).  MSPs are formed through accretion from a companion
during an X-ray binary phase~{\bfref \citep{1982Natur.300..728A,1991PhR...203....1B}}
in which the pulsar is ``recycled''. This accretion phase spins the
pulsar up to very fast rotational rates (spin periods $P \lesssim30$
ms). Such pulsars are useful for a variety of physical applications.
Examples include tests of theories of gravity using MSP-white dwarf systems 
such as PSR J1738+0333 and PSR J0348+0432~\citep{2012MNRAS.423.3328F,2013Sci...340..448A}
and triple systems like PSR J0337+1715~\citep{2014Natur.505..520R};
{\bfref tests of General Relativity using double neutron star systems, such as
J0737$-$3039~\citep{2006Sci...314...97K} and PSR B1913+16~\citep{2010ApJ...722.1030W};}
the study of binary systems such as eccentric MSPs like PSRs
J1903+0327~\citep{2008Sci...320.1309C} and J1950+2414~\citep{2015ApJ...806..140K} {\bfref
which are interesting due to their peculiar binary evolution};
and constraining the equation-of-state of dense matter using measurements
of neutron star masses~\citep{2010Natur.467.1081D,2013Sci...340..448A}. Another major
driver for the discovery of new
MSPs is the effort to detect gravitational wave emission using an
array of pulsars~\citep{2015ApJ...813...65T,
  2015MNRAS.453.2576L,2016MNRAS.455.1751R}.  The large-scale pulsar
surveys mentioned above, combined with targeted searches of unidentified gamma-ray
sources from the \textit{Fermi Gamma-Ray Space Telescope}~\citep[e.g.][]{
2011AIPC.1357...40H,2011ApJ...727L..16R,2011MNRAS.414.1292K,
2012ApJ...748L...2K}, have resulted in the discovery of about 90
new MSPs in the past 5 years, an increase of 40\%
in the known Galactic MSP population.  A subset of the newly discovered sources
are eclipsing systems that appear to fall into two
categories~\citep[e.g.][]{2005ASPC..328..405F,2011AIPC.1357..127R}. The first category, known as
black widow systems, has
very-low-mass{\bfref, degenerate} companions ($\lesssim0.05\,M_\mathrm{\odot}$) {\bfref believed
to be the result of ablation by the pulsar.}
{\bfref T}he
second, known as redback systems, has low{\bfref-} to moderate{\bfref-}mass{\bfref, non-degenerate} companions
($M\sim0.15-0.7\,M_\mathrm{\odot}$).

The PALFA survey~\citep{2006ApJ...637..446C,2015ApJ...812...81L} is an ongoing search for
new pulsars and transients in the Galactic plane ($|b|<5^{\circ}$)
that is accessible to the Arecibo Observatory William E. Gordon 305-m
Telescope using the ALFA 7-beam receiver.  The survey consists of an inner-Galaxy region ($32^{\circ}
\lesssim l \lesssim 77^{\circ}$) and an outer Galaxy region
($168^{\circ} \lesssim l \lesssim 214^{\circ}$). The relatively high
observing frequency used in the survey (1.4 GHz) mitigates the
deleterious effects present from the ISM that can prevent detection of
rapidly spinning pulsars. This makes the PALFA survey well suited to
discovering highly dispersed (distant) MSPs in the Galactic plane.
Hence, PALFA is providing a view of the Galactic MSP
population that complements what is being found at high Galactic
latitudes by all-sky and targeted searches.
PALFA began in 2004 and to date has discovered 165 radio
pulsars~\citep{2006ApJ...637..446C,2013ApJ...772...50N,2015ApJ...812...81L},
including 26 MSPs~\citep[e.g.][]{2008Sci...320.1309C,2010Sci...329.1305K,2011ApJ...732L...1K,2012ApJ...757...90C,2012ApJ...757...89D,2013ApJ...773...91A,2015ApJ...800..123S,2015ApJ...806..140K} and 1 repeating fast radio burst~\citep{2014ApJ...790..101S,2016Natur.531..202S}.

Here we present the discovery and follow-up timing of five MSPs found in the PALFA
survey: PSRs J1906$+$0055, J1914$+$0659, J1933$+$1726, J1938$+$2012,
and J1957$+$2516.  In Section~\ref{sec:obs}, we describe the
observations used to discover and time these systems. In
Section~\ref{sec:disc}, we describe the details of each pulsar
system. In Section~\ref{sec:conc}, we present our conclusions.

\section{Observations and Analysis}\label{sec:obs}

\subsection{Discovery}\label{subsec:disc}
The MSPs described here were discovered in the inner-Galaxy region of
the PALFA survey between 2010 September and 2012 September. During this
time, the PALFA survey used the Mock spectrometers\footnote{http://www.naic.edu/{\tt\~{}}astro/mock.shtml} to record data from
the 7-beam ALFA receiver centered at 1375 MHz with 322.617 MHz of
bandwidth {\bfref across 960 channels} and a sample time of 65.476 $\mathrm{\mu s}$. 
PALFA uses 268-s integrations per pointing in the inner-Galaxy.
Additional details can be found in~\cite{2015ApJ...812...81L}.

The PALFA survey uses 3 pipelines to search for radio pulsars: (1) a
full-resolution {\tt PRESTO\footnote{http://www.cv.nrao.edu/{\tt\~{}}sransom/presto/}}-based  pipeline~\citep{2015ApJ...812...81L},
(2) the Einstein@Home pulsar search pipeline~\citep{2013ApJ...773...91A},
and (3) a {\tt PRESTO}-based reduced-time-resolution ``Quicklook''
pipeline~\citep{2013PhDT.......127S}. The 5 MSPs presented here were
all discovered using pipeline (1) on the Guillimin supercomputer operated
for Compute Canada by McGill University, so it is the only pipeline we
describe in {\bfref some} detail.

{\bfref Prior to searching the data for pulsars,
pipeline (1) performs radio frequency interference (RFI) excision which consists
of multiple components. The first removes known RFI that is specific to electronics
at the Arecibo Observatory. Then narrow-band RFI is identified in both the time- and
frequency-domain using {\tt PRESTO}'s {\tt rfifind} routine. The narrow-band RFI is
masked out of subsequent processing. Next, broadband signals are identified by analyzing
the DM=0 $\mathrm{pc\;cm^{-3}}$ timeseries. Bad time intervals are identified as samples
whose values are more than 6 times larger than the local standard deviation. Such
samples are replaced with the local median bandpass. At this stage, the data is de-dispersed
into trial dispersion measures (DMs) ranging from 0 to $\sim$10,000 $\mathrm{pc\;cm^{-3}}$.
The DM steps are chosen such that the maximum DM smearing is about 0.1 ms at low DMs, increases
to 1 ms for DMs of a few hundred $\mathrm{pc\; cm^{-3}}$, and reaches 10 ms at the highest DMs.
After de-dispersion, two additional RFI excision steps are performed. The first is to remove ``red''
noise from each de-dispersed time-series using {\tt PRESTO}'s {\tt rednoise} routine which performs
a median median removal using logarithmically-increasing block sizes in the frequency domain. The
second is to remove Fourier bins from the power spectrum of RFI identified in lists of identified
RFI that are generated dynamically from the combined ALFA beams and for each beam from the entire
observing session.} Each de-dispersed
timeseries is searched for periodic signals using common Fourier search
techniques and for individual pulses in the time domain using the
{\tt PRESTO} software
package~\citep{2002AJ....124.1788R}.
The periodic signal search has two components: the first is a zero-acceleration
search and the second searches for signals with constant accelerations up
to $\sim\mathrm{1650\;m\;s^{-2}}$~\citep{2015ApJ...812...81L}. The best
candidates (roughly 100 per search pointing) are folded into diagnostic
plots for further evaluation. Due to the large number of candidates that
are generated by this pipeline, we have investigated multiple ways of
sorting through them. One method uses machine learning
algorithms to sort through the candidates using image pattern
recognition~\citep{2014ApJ...781..117Z}. PSR J1938$+$2012 was identified
using this image pattern recognition technique, while the other 4 MSPs
were identified by sorting on various heuristic ratings~(see Table~4 of \cite{2015ApJ...812...81L}).

\subsection{Timing Observations}\label{subsec:timobs}
After discovery, follow-up observations of each of the 5 new MSPs were conducted
in order to determine rotational, astrometric, and binary parameters
where applicable. This is done by accounting for every rotation of the pulsar
over the entire data span by observing at appropriate spacings in time such that
the number of pulses between observations is unambiguous. Once we obtained phase-connected
solutions spanning more than one month, we observed each pulsar on a roughly monthly
basis. PSRs J1906$+$0055, J1933$+$1726, J1938$+$2012, and J1957$+$2516
were timed using the 305-m William G. Gordon Telescope at the Arecibo Observatory while
PSR J1914$+$0659 was timed using the 76-m Lovell Telescope at the Jodrell Bank Observatory.

From 2011 September 9 to 2013 September 20, observations of PSRs J1906$+$0055 and J1957$+$2516
at the Arecibo Observatory were performed using the ALFA receiver with the Mock spectrometers.
These observations were typically conducted as test pulsars for the PALFA survey and were
therefore recorded in the same mode as search observations described in~\cite{2015ApJ...812...81L}
and had integration times of between 268 and 600 seconds.

After 2012 February 18, follow-up observations of PSRs J1906$+$0055, J1933$+$1726, J1938$+$2012,
and J1957$+$2516 were conducted using the L-wide receiver with the Puerto Rican Ultimate Pulsar
Processing Instrument (PUPPI), which is a clone of the {\bfref Green Bank} Ultimate Pulsar Processing
Instrument (GUPPI)\footnote{https://safe.nrao.edu/wiki/bin/view/CICADA/GUPPiUsersGuide}. Initial observations consisted of incoherent search mode observations
with 800 MHz of bandwidth that was split into 2048 channels with a sample time of 40.96 $\mu s$. 
These data were then folded using the {\tt fold\_psrfits} routine
from the {\tt psrfits\_utils} software package\footnote{http://github.com/scottransom/psrfits\_utils}.
Later observations were performed using PUPPI in coherent fold mode with the same 800 MHz of
bandwidth split into 512 frequency channels and were written to disk every 10 s.
RFI was excised from both incoherent and coherent PUPPI files using a median-zapping algorithm
included in the {\tt PSRCHIVE{\bfref \footnote{http://psrchive.sourceforge.net/}}} software package\citep{2012AR&T....9..237V}.

Follow-up timing observations of PSR J1914$+$0659 at Jodrell Bank were done using a dual polarization cryogenic receiver
with a center frequency of 1525 MHz and bandwidth of 350 MHz. Data were processed by a digital
filter bank producing 700 frequency channels of 500 kHz bandwidth. The output of each channel was folded
at the nominal topocentric period of the pulsar and the resultant profiles written to disk every 10 s.
RFI was removed using a median-zapping algorithm and the data were then incoherently de-dispersed
at the pulsar's DM and folded profiles were produced for total integration times of typically 40 minutes.

\subsection{Timing Analysis}\label{subsec:timan}
\begin{figure}[h!]
\centering
%\begin{center}
\includegraphics[width=0.5\textwidth]{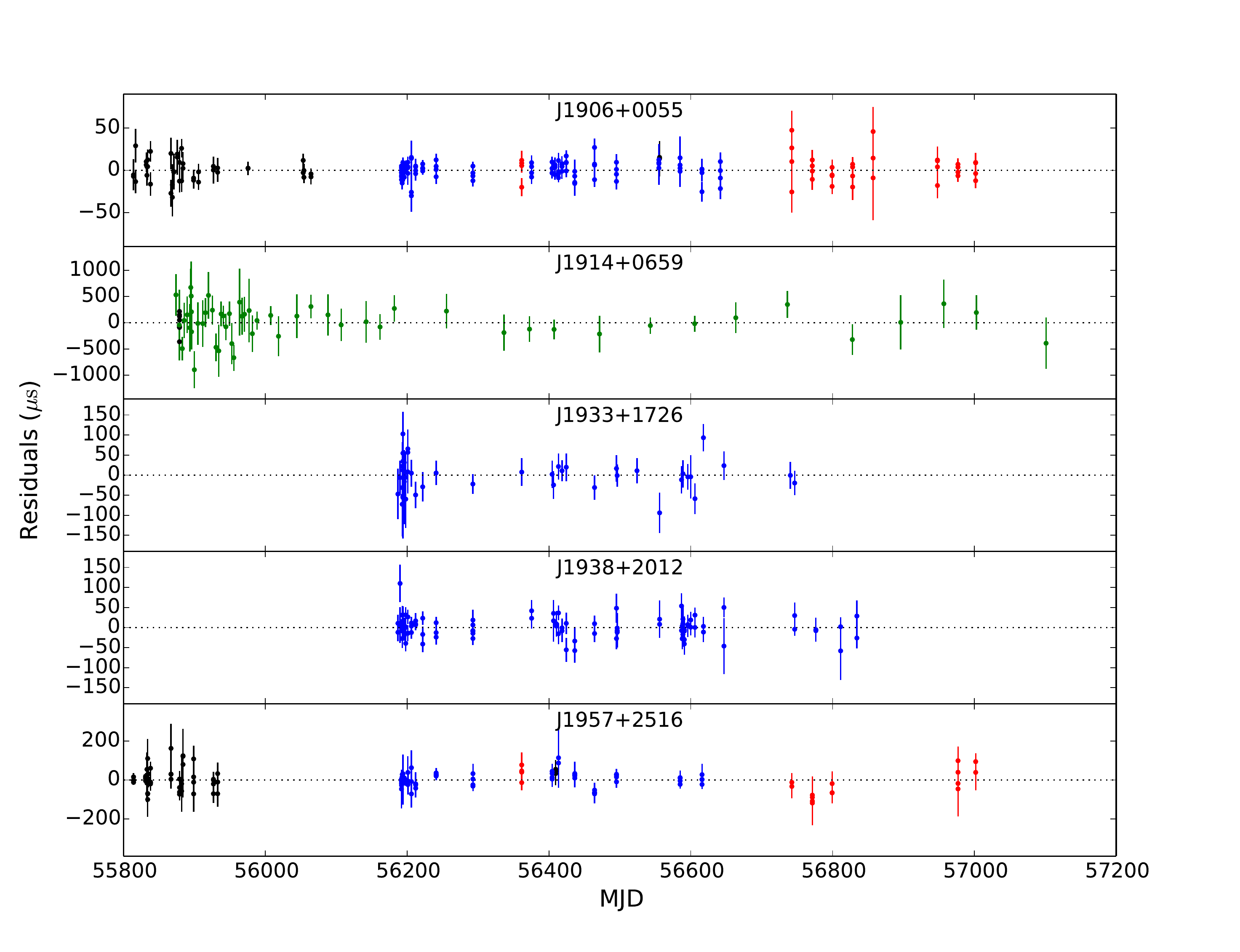}
\caption[Residuals for 5 PALFA MSPs]{Post-fit timing residuals versus MJD for 5 MSPs discovered in the
PALFA survey. Green points are from Jodrell Bank observations, black points are from Arecibo Mock spectrometer
observations, blue points are from Arecibo PUPPI observations in incoherent search mode, and
red points are from PUPPI observations in coherent fold mode.}
\label{fig:resids}
%\end{center}
\end{figure}

In order to construct the initial timing solutions for the 4 binary pulsars, we performed a fit
of the observed periods from early observations to orbital Doppler shifts for circular orbits.
For each of the 5 pulsars, we constructed pulse templates by summing together data from multiple observations
and fitting Gaussian components to the summed profiles. Examples of summed profiles for each of the
pulsars presented here were previously included
in~\cite{2015ApJ...812...81L}. For each observation, we generated pulse times-of-arrival (TOAs) using
{\tt pat} from the {\tt PSRCHIVE} software package to
cross-correlate the pulsar's template profile with the data in the Fourier domain~\citep{1992PTRSL.341..117T}.
The observations from Arecibo were split into 2 to 4 frequency sub-bands depending on the pulsar's
brightness, and each sub-band was used to generate a TOA allowing us to fit for the DM. 
The TOAs were then fitted to a model for each pulsar to get the final timing solutions using
{\tt TEMPO}\footnote{http://tempo.sourceforge.net} {\bfref For all timing solutions, we used the
DE421 solar system ephemeris and the BIPM clock timescale}. The final timing solutions are given in
Table~\ref{table:solutions} and the timing residuals are shown in Figure~\ref{fig:resids}.

\begin{figure}[h!]
\centering
%\begin{center}
\includegraphics[width=0.5\textwidth]{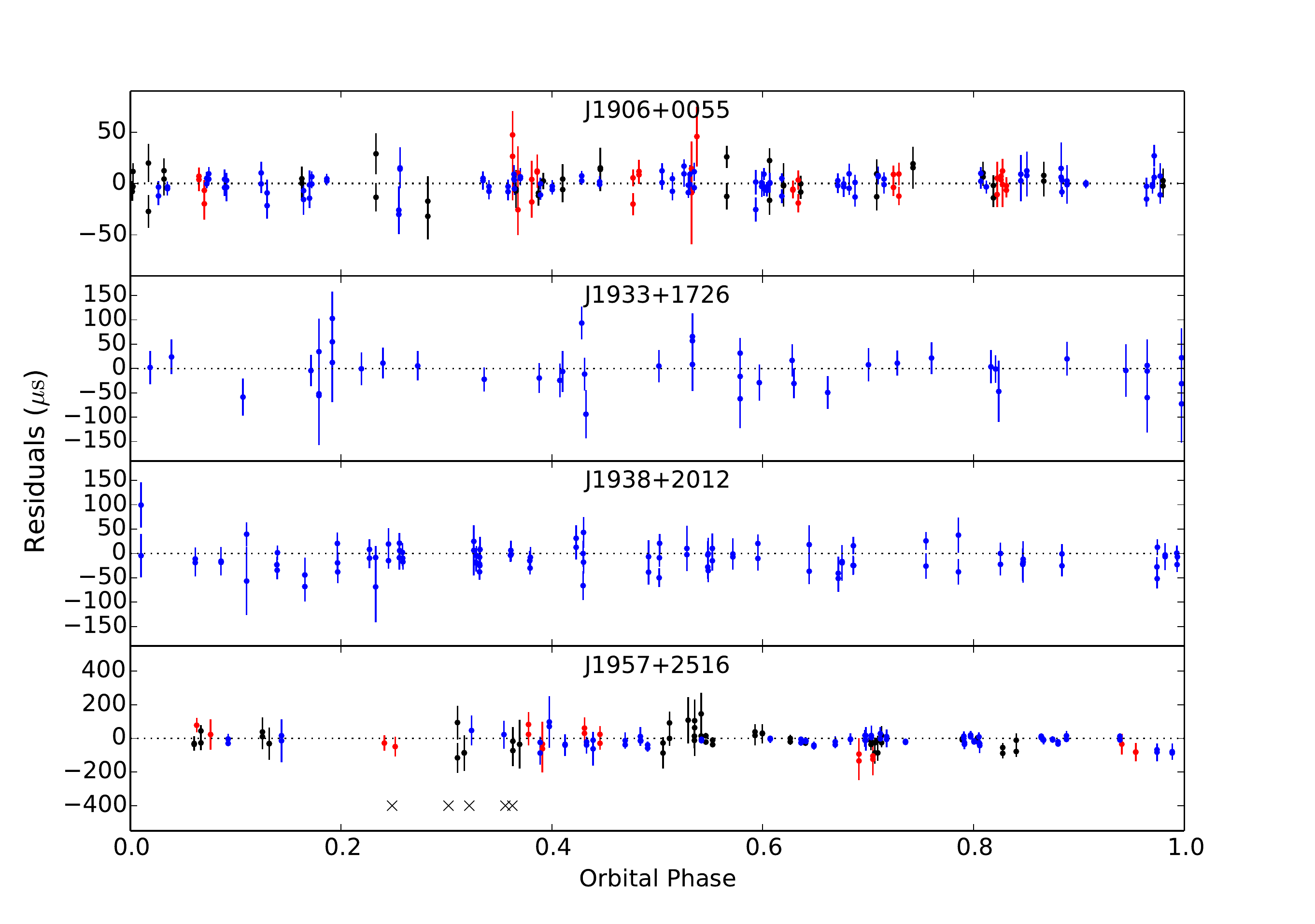}
\caption[Orbital Residuals for 5 PALFA MSPs]{Post-fit timing residuals versus orbital phase for 4 binary
 MSPs discovered in the PALFA survey. PSR J1957$+$2516 is often eclipsed between orbital phases
 0.15 and 0.4. Non-detections with observing times of at least 600-s are shown as black X's at the bottom
of the plot. However, there are also detections within this orbital phase range. Black points
are from Arecibo Mock spectrometer
observations, blue points are from Arecibo PUPPI observations in incoherent search mode, and
red points are from PUPPI observations in coherent fold mode. }
\label{fig:orbresids}
%\end{center}
\end{figure}

All of the binary pulsars presented here are in highly circular orbits (eccentricity $<$ $10^{-4}$). Since
there is a strong
correlation between the longitude of periastron ($\omega$) and the epoch of periastron passage
($T_{0}$) in circular binaries, we used the ELL1 binary model in {\tt TEMPO} (see the Appendix
of \citealt{2001MNRAS.326..274L}). The ELL1 model removes this correlation by using
the parameterization $\epsilon_1=e\;\sin\;\omega$, $\epsilon_1=e\;\cos\;\omega$, and
$T_{asc} = T_{0}-\omega P_b / 2\pi$ where $e$ is the eccentricity and $P_b$ is the orbital
period. This parameterization is superior for cases where $x e^{2}$
($x$ is the projected semi-major axis) is much smaller than the error in the TOAs, as is the
case for all pulsars presented here. The timing residuals as a function of orbital period
are presented in Figure~\ref{fig:orbresids}.

The procedure described in Section~\ref{subsec:timan} is known to often result in the underestimation of TOA errors.
Therefore, the TOA uncertainties for each recording instrument mode were multiplied by a scaling factor
to produce a fit with $\chi^2$ equal to 1 for that subset of residuals. This results in more conservative estimates
for the timing parameter uncertainties. The scaling factors (EFACs) used for each pulsar are given in
Table~\ref{table:solutions}.

\clearpage
\begin{turnpage}
%\begin{landscape}
\begin{table}[h]
\caption{Timing solutions and derived parameters for 5 PALFA discovered MSPs.}
  \begin{center} {\footnotesize
  \begin{tabular}{lccccc}
  Parameter &
   PSR J1906+0055 &
   PSR J1914+0659 &
   PSR J1933+1726 &
   PSR J1938+2012 &
   PSR J1957+2516 \\
    \hline
    \multicolumn{6}{c}{Timing Parameters} \\
    \hline\hline
    Right Ascension (J2000)                       \dotfill & 19:06:48.68051(4)             & 19:14:17.647(2)               & 19:33:22.9828(3)              & 19:38:40.0803(1)              & 19:57:34.6115(3)              \\
    Declination (J2000)                           \dotfill & 00:55:07.886(1)               & 07:01:11.00(7)                & 17:26:49.606(9)               & 20:12:50.827(3)               & 25:16:02.076(3)               \\
    Pulsar Period ($\s$)                          \dotfill & 0.0027895524236884(2)         & 0.01851182255144(3)           & 0.02150723378644(1)           & 0.0026341351275486(6)         & 0.003961655342404(1)          \\
    Period Derivative ($\mathrm{s}\, \ps$)                 \dotfill & 3.32(1)$\times$10$^{-21}$     & 3.1(3)$\times$10$^{-20}$      & 4.9(1)$\times$10$^{-20}$      & 7.5(6)$\times$10$^{-22}$      & 2.744(9)$\times$10$^{-20}$    \\
    Dispersion Measure (\dmu)                     \dotfill & 126.8317(9)                   & 225.3(2)                      & 156.90(3)                     & 236.909(5)                    & 44.137(3)                     \\
    Reference Epoch (MJD)                         \dotfill & 56408.0                       & 56351.0                       & 56466.0                       & 56511.0                       & 56408.0                       \\
    Span of Timing Data (MJD)                     \dotfill & 55814--57001                  & 55873--57101                  & 56186--56746                  & 56186--56834                  & 55814--57001                  \\
    Number of TOAs                                \dotfill & 187                           & 61                            & 45                            & 96                            & 151                           \\
    RMS Residual ($\us$)                          \dotfill & 7.84                          & 245.61                        & 33.95                         & 20.19                         & 19.15                         \\
    EFAC (JB/Mock/PI/PC)$^a$                          \dotfill & -/1.0/1.0/1.4               & 1.0/1.1/-/-               & -/-/1.1/-               & -/-/1.1/-               & -/1.5/2.0/2.1               \\
    1400 MHz mean flux density (mJy)              \dotfill & 0.1                       & \nodata & 0.04 & \nodata & 0.02 \\
    \hline
    \multicolumn{6}{c}{Binary Parameters} \\
    \hline\hline
    Orbital Period (days)                         \dotfill & 0.6096071304(3)               & \nodata                       & 5.15393626(2)                 & 16.2558195(1)                 & 0.2381447210(7)               \\
    Orb. Per. Derivative ($\mathrm{s}\, \ps$)              \dotfill & \nodata & \nodata & \nodata & \nodata & 12(2)$\times$10$^{-12}$ \\
    Projected Semi-major Axis (lt-s)              \dotfill & 0.6250279(9)                  & \nodata                       & 13.67353(1)                   & 8.317778(4)                   & 0.283349(6)                   \\
    Epoch of Ascending Node (MJD)                 \dotfill & 56407.5586451(1)              & \nodata                       & 56466.1820553(6)              & 56514.938959(1)               & 56407.8681189(6)              \\
    ECCsin(OM)                                    \dotfill & -1.0(3)$\times$10$^{-6}$      & \nodata                       & -1.8(1)$\times$10$^{-5}$      & 9.9(8)$\times$10$^{-6}$       & 2(3)$\times$10$^{-5}$         \\
    ECCcos(OM)                                    \dotfill & 1(2)$\times$10$^{-6}$         & \nodata                       & 6.5(1)$\times$10$^{-5}$       & -3.2(9)$\times$10$^{-6}$      & 2(2)$\times$10$^{-5}$         \\
    Mass Function (\Msun)                         \dotfill & 0.0007055                     & \nodata                       & 0.1033                        & 0.002338                      & 0.0004307                     \\
    Min. Companion Mass (\Msun)                   \dotfill & 0.12                          & \nodata                       & 0.79                          & 0.18                          & 0.099                         \\
    Med. Companion Mass (\Msun)                   \dotfill & 0.14                          & \nodata                       & 0.96                          & 0.21                          & 0.12                          \\
    \hline
    \multicolumn{6}{c}{Derived Parameters} \\
    \hline\hline
    Galactic Longitude ($^\circ$)                 \dotfill & 35.51                         & 41.79                         & 53.18                         & 56.2                          & 62.77                         \\
    Galactic Latitude  ($^\circ$)                 \dotfill & -3.0                          & -1.85                         & -1.02                         & -0.76                         & -1.97                         \\
    Eccentricity                                  \dotfill & 1.4$\times$10$^{-6}$          & \nodata                       & 6.7$\times$10$^{-5}$          & 9.4$\times$10$^{-6}$          & 2.8$\times$10$^{-6}$          \\
    DM Derived Distance$^b$ (kpc)                     \dotfill & 3.3                           & 6.1                           & 5.5                           & 7.7                           & 3.1                           \\
    Surface Mag. Field Strength (G)               \dotfill & 0.97$\times 10^{8}$           & 7.7$\times 10^{8}$           & 10.4$\times 10^{8}$           & 0.45$\times 10^{8}$         & 3.3$\times 10^{8}$           \\
    Spindown Luminosity (erg/s)                   \dotfill & 60.4$\times 10^{32}$          & 1.96$\times 10^{32}$          & 1.95$\times 10^{32}$          & 16.2$\times 10^{32}$          & 174.0$\times 10^{32}$         \\
    Characteristic Age (Gyr)                      \dotfill & 13.3                          & 9.34                          & 6.92                          & 55.6                          & 2.29                          \\
  \footnotetext[0]{Numbers in parentheses represent 1-$\sigma$ uncertainties in the
last digits as determined by \texttt{TEMPO}, scaled such that the reduced
$\chi^2 = 1$.  All timing solutions use the DE421 Solar System Ephemeris and the
UTC(BIPM) time system.   Derived quantities assume an $R = 10\; \km$ neutron star with
$I = 10^{45}\; \gm\, \cm^2$.  Minimum companion masses were calculated assuming a
$1.4\; \Msun$ pulsar. The EFAC values correspond to subsets of TOAs from observations from Jodrell Bank (JB), Mock spectrometers and the ALFA receiver (Mock),
PUPPI with L-wide in incoherent search mode (PI), and PUPPI with L-wide in coherent search mode (PC). The \DM\ derived distances were calculated using the NE2001
model of Galactic free electron density, and have typical errors of $\sim 20\%$
\citep{2002astro.ph..7156C}.}
\end{tabular} }
\end{center}
%\vspace{-1em}
%\caption{Timing solutions and derived parameters for 5 PALFA discovered MSPs.}
%\vspace{-1em}
\label{table:solutions}
\end{table}
%\end{landscape}
\end{turnpage}
\clearpage

\subsection{Mean Flux Density Measurements}
For the PUPPI observations taken in coherent mode we also obtained observations of a noise
diode, which are suitable for use in polarization calibration. Such observations were
made for PSRs J1906$+$0055, J1933$+$1726, and J1957$+$2516. We used observations
of the bright quasar B1442+10 made by the NANOGrav Collaboration~\citep{2015ApJ...813...65T} for
flux calibration. The calibration of polarization and flux were performed using the {\tt pac} tool
from the {\tt PSRCHIVE} analysis package using the {\it SingleAxis} model. We searched for rotation
measures in the range -2,000 to 2,000 $\mathrm{rad\;m^{-2}}$, but did not detect substantial 
polarization for any of these 3 MSPs. The mean flux densities are included in Table~\ref{table:solutions}.

\section{Discussion}\label{sec:disc}
We have reported the timing solutions for 5 MSPs discovered in
the PALFA survey. These MSPs consist of 4 circular binary systems and one isolated system.
The DMs of these MSPs range from 44 to to 236 $\mathrm{pc\;cm^{-3}}$. Two of them
have DM/P ratios greater than 30 $\mathrm{pc\;cm^{-3}\;ms^{-1}}$ (see Fig.~\ref{fig:dmvp0}), adding to the population
of high-DM MSPs being discovered by the PALFA survey~\citep{2012ApJ...757...90C,2015ApJ...800..123S}. {\bfref The
eccentricities of the 4 circular binaries are near the expected values for their respective orbital periods from
the relation in~\cite{1992RSPTA.341...39P} indicating these systems have been circularized through mass transfer from
the companion. Also, the 3 low mass systems are all consistent with the expected companion mass--orbital period relation
for HeWD-MSP systems formed through long-term mass transfer in low-mass X-ray binaries~\cite[LMXBs;][]{1999A&A...350..928T}} Below, we discuss details of each individual MSP.

\subsection{PSR J1906$+$0055}\label{subsec:1906}
PSR J1906$+$0055 is a 2.8-ms pulsar with a DM of 127 $\mathrm{pc\;cm^{-3}}$. It is in a 
circular, 14.6-hour orbit with a companion having a minimum mass of 0.12 $M_{\odot}$ (assuming a
pulsar mass of 1.4 $M_{\odot}$
and an inclination angle of $90^{\circ}$) and a median mass of 0.14 $M_{\odot}$ (assuming
a pulsar mass of 1.4 $M_{\odot}$ and an inclination angle of $60^{\circ}$). Therefore, the 
companion is likely a He white dwarf. The DM-derived distance for J1906$+$0055 from the NE2001
model~\citep{2002astro.ph..7156C} for the Galactic electron density is 3.3 kpc. PSR J1906$+$0055's
14.6-hour orbit makes it a potential candidate for detection of orbital decay and for constraining
dipolar gravitational radiation~\citep{2012MNRAS.423.3328F}.

\subsection{PSR J1914$+$0659}\label{subsec:1914}
PSR J1914$+$0659 is an isolated, partially recycled pulsar with a spin period of 18.5 ms
and a DM of 225 $\mathrm{pc\;cm^{-3}}$. It's DM-derived distance is 6.1 kpc. Based on the
spin period and period derivative, it is likely the result of a disrupted double neutron
star system~{\bfref\citep{2004MNRAS.347L..21L}}.

\subsection{PSR J1933$+$1726}\label{subsec:1933}
PSR J1933$+$1726 is a partially recycled pulsar with a spin period of 22 ms and a DM of 
157 $\mathrm{pc\;cm^{-3}}$ in a circular, 5-day orbit with a fairly high-mass companion. The 
companion's minimum mass is 0.8 $M_{\odot}$. This companion mass combined with the low orbital
eccentricity
indicates the companion is likely to be a CO white dwarf{\bfref ~\citep{2001ApJ...548L.187C}}. The DM-derived distance for
this system is 5.5 kpc. The relatively high companion mass makes this system a candidate
for a future Shapiro delay measurement if it has a favorable orbital inclination angle.

\subsection{PSR J1938$+$2012}\label{subsec:1938}
\begin{figure}[h!]
\centering
\includegraphics[width=0.5\textwidth]{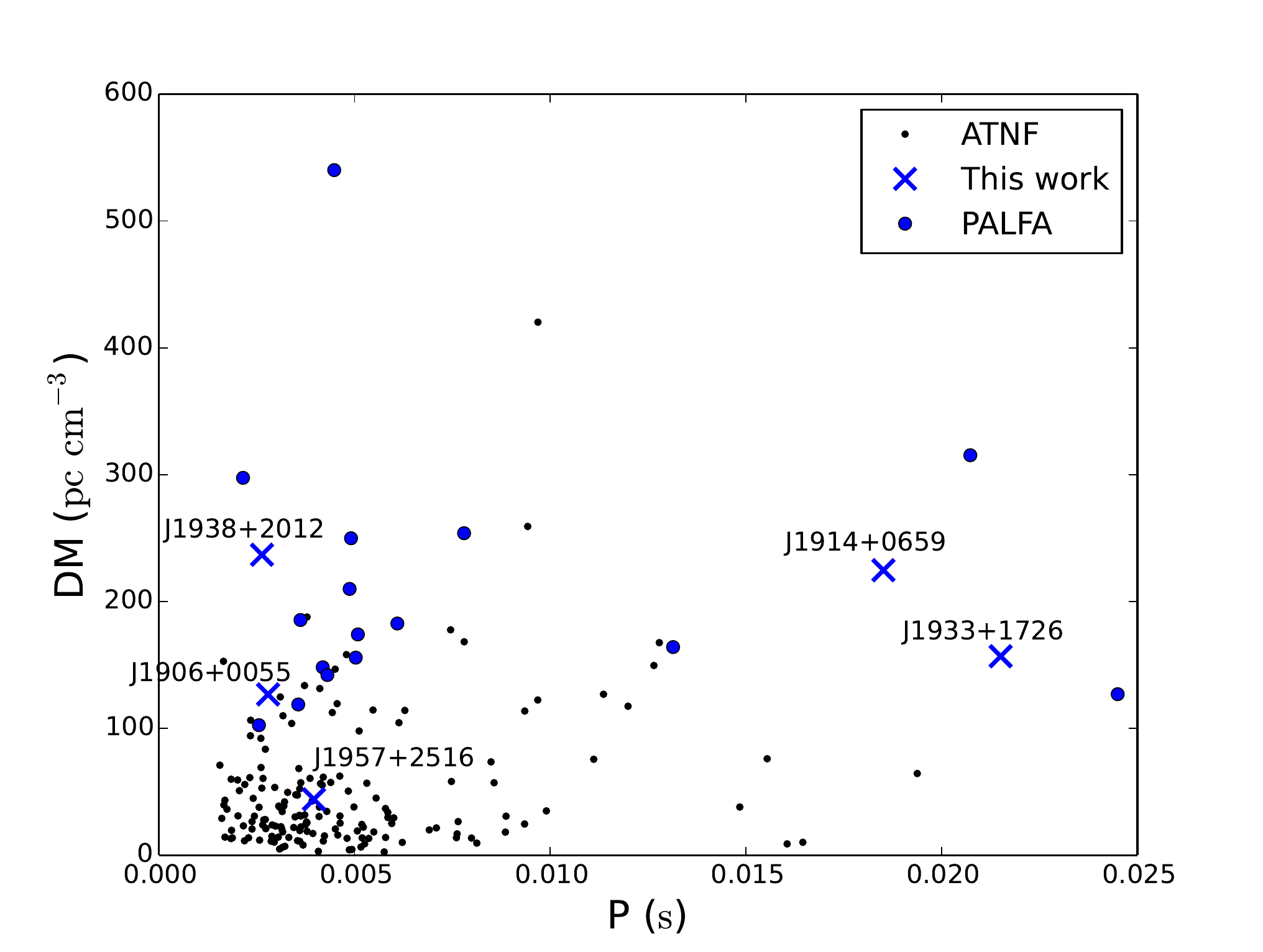}
\caption[DM vs Spin Period for 5 PALFA MSPs]{
Spin period versus DM for the 5 MSPs presented here are shown
along with known Galactic MSPs (cf. Figure 5 of~\citealt{2012ApJ...757...90C} and 
Figure 1 of~\citealt{2015ApJ...800..123S}). The Galactic MSPs are from the ATNF pulsar catalog~\citep{2005AJ....129.1993M}}.
\label{fig:dmvp0}
\end{figure}

PSR J1938$+$2012 is a 2.6-ms pulsar with a DM of 237 $\mathrm{pc\;cm^{-3}}$ and is in a circular, 16-day 
orbit. The companion's minimum mass is 0.2 $M_{\odot}$ and is therefore likely a He white 
dwarf. Its DM is among the highest known for rapidly-rotating MSPs (P$\le$5 ms; see Fig.~\ref{fig:dmvp0})
The DM-derived distance for this system is 7.7 kpc.

\subsection{PSR J1957$+$2516}\label{subsec:1957}

\begin{figure}[h!]
\centering
\includegraphics[width=0.5\textwidth]{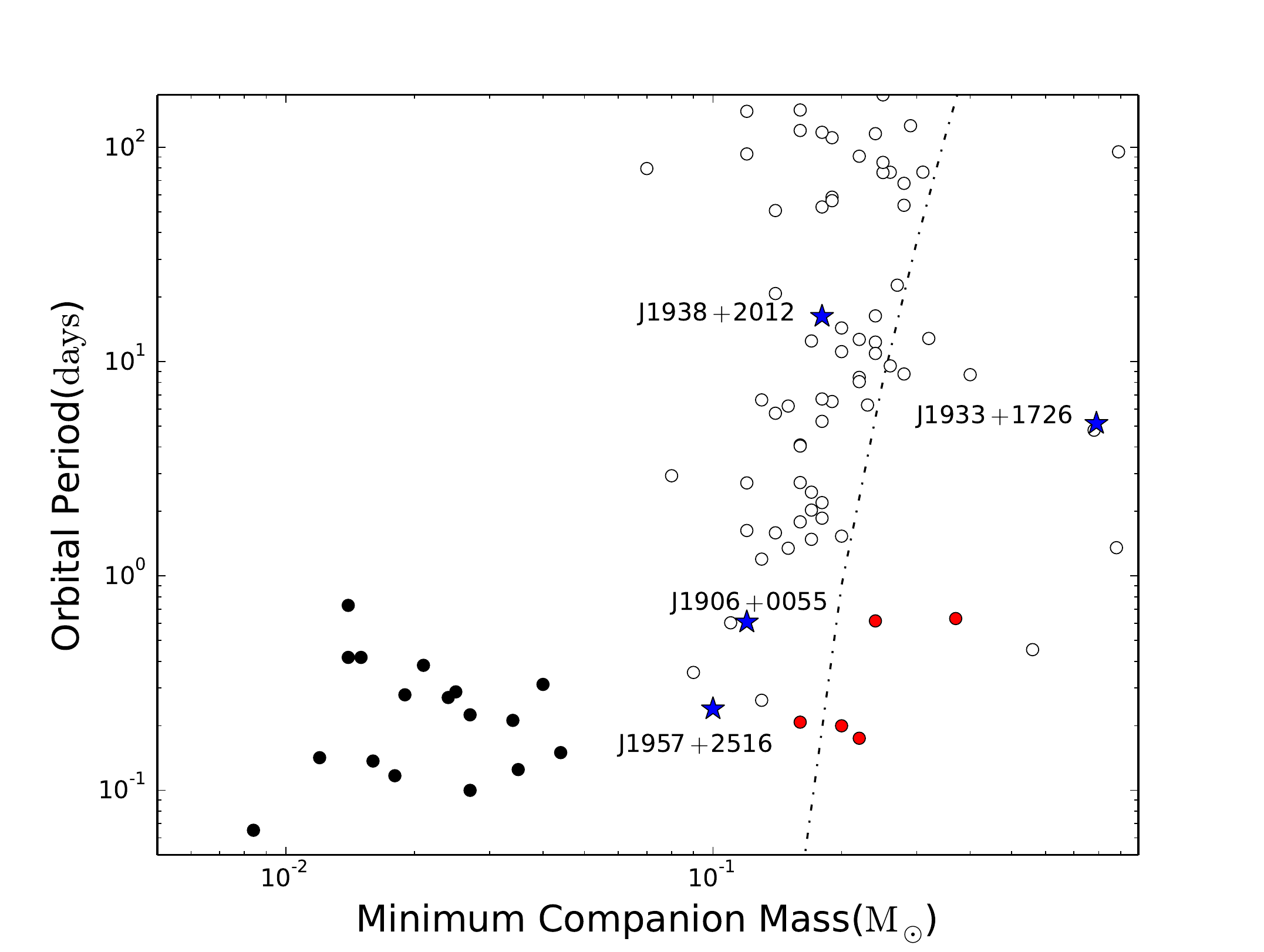}
\caption[Mass vs Orbital Period for 4 PALFA MSPs]{Orbital period versus minimum companion mass is shown 
for the 4 binary PALFA MSPs presented here (blue stars) and known Galactic field MSPs.
Plotted are MSPs from the ATNF pulsar catalog~\citep[unfilled black circles;][]{2005AJ....129.1993M}, 
known redbacks (filled
red circles), and known black widows (filled black circles). The dashed line
shows the relation between companion mass and orbital period {\bfref for systems formed through long-term
mass transfer in LMXBs} from~\cite{1999A&A...350..928T}. {\bfref As expected, the 3 MSPs with low mass
companions are near the relation.}}
\label{fig:massvpb}
\end{figure}

PSR J1957$+$2516 is a 4.0-ms pulsar with a DM of 44 $\mathrm{pc\;cm^{-3}}$ that is in a circular, 6.8-hour
orbit. We have plotted the orbital phase of non-detections of PSR J1957$+$2516 during observations
of 600-s or longer in Fig.~\ref{fig:orbresids} as black X's. These non-detections are consistent
with eclipsing as they occur near the orbital phase of 0.25 which corresponds to superior conjunction.
However, we cannot rule out that these non-detections are not due to scintillation, since the pulsar
has a fairly low DM. There have also been detections of the pulsar during portions of the orbit where
it appears to have been eclipsed at other times (see Fig.~\ref{fig:orbresids}). However, we note that the
detections that occured near an orbital phase of 0.25 are from the top of the band only (i.e. the pulsar
was not detected below about 1.4 GHz in that session). In addition to the possible eclipses, we also 
detected an orbital period derivative of
12$\times10^{-12}$, making this a likely redback or black widow system.
PSR J1957$+$2516's companion mass is $\approx$0.1 $M_{\odot}$ and is consistent
with this being a redback system, however as shown in Fig.~\ref{fig:massvpb} it lies between the
black widow and redback populations in the orbital period versus minimum companion mass diagram.

\subsection{Counterparts at Other Wavelengths}\label{subsec:counterparts}
\begin{figure}[h!]
\centering
\includegraphics[width=0.5\textwidth]{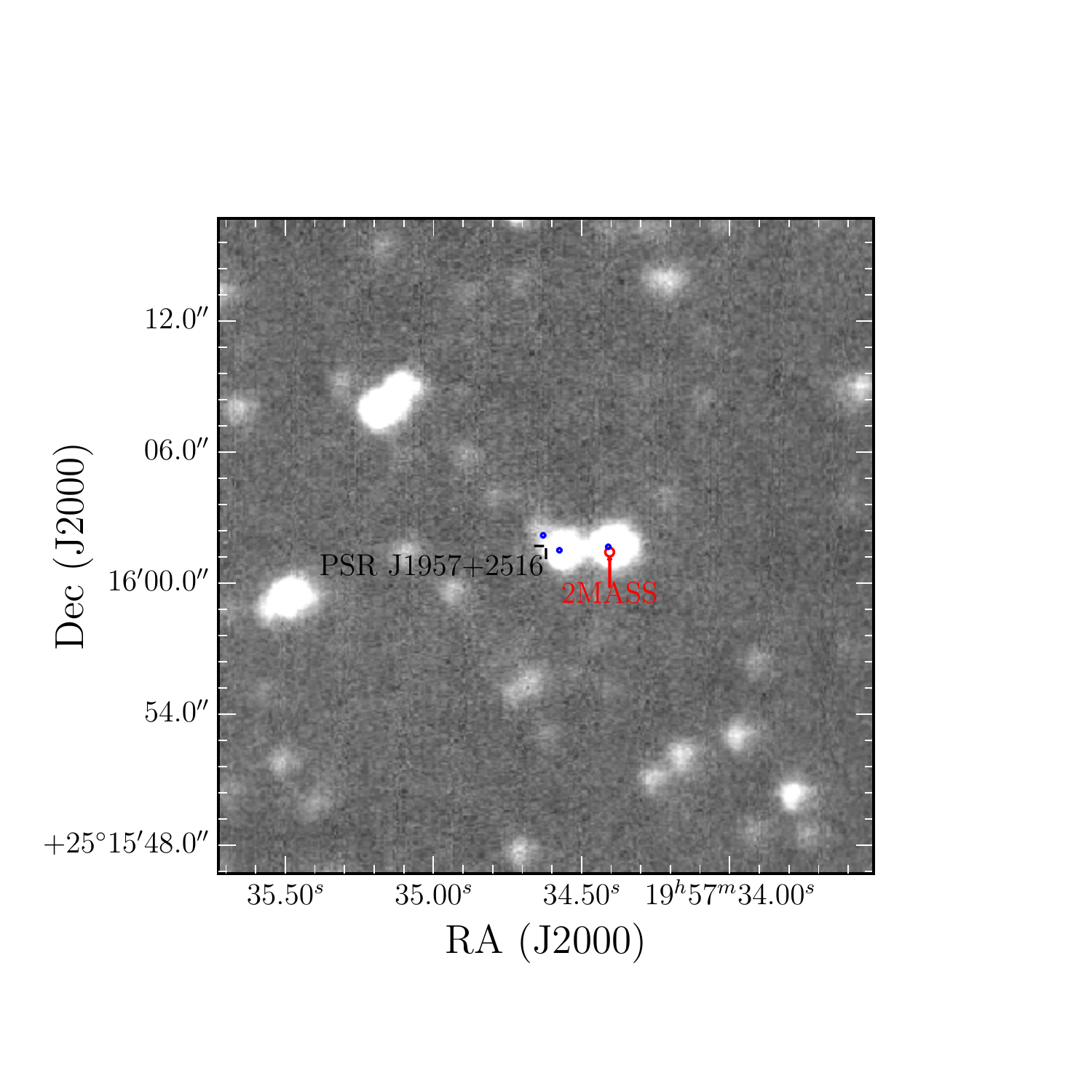}
\caption[IR Observation of PSR J1957$+$2516]{WIYN 3.5-m observation of the field
containing PSR J1957$+$2516. The position obtained from the radio timing of PSR J1957+2516 is shown using black tick marks. The positions of the 3 stars described in Section~\ref{subsec:counterparts} are shown using blue circles. }
\label{fig:ir}
\end{figure}

For all of the binary pulsars presented here, we examined archival optical and
infrared data\footnote{http://irsa.ipac.caltech.edu} for possible counterparts
at the locations give in Table~\ref{table:solutions}. 
Only PSR J1957$+$2516 was found to have a potential counterpart.
We identified a potential near-infrared counterpart to PSR J1957$+$2516 from the Two Micron 
All-Sky Survey (2MASS; \citealt{2006AJ....131.1163S}). 2MASS~J19573440+2516014
lies $2\farcs5$ away from the pulsar.  This is considerably larger than the typical {\bfref 2MASS
astrometric uncertainty of $\sim0\farcs1$}, but it led us to further examine the field.  On
visually inspecting the 2MASS image the source appeared extended in the east-west
direction {\bfref by $\sim5''$, overlapping with the position of PSR J1957+2516}. Therefore
we obtained further near-infrared imaging {\bfref with the} WIYN High Resolution
Infrared Camera (WHIRC; \citealt{2010PASP..122..451M}) on the WIYN 3.5-m telescope {\bfref in order
to resolve the extended source}. 
Our data consist of $14\times 30\,$s exposures with the
$J$ filter and $11\times 30\,$s exposures with the $K_s$ filter on the night of
2014~May~10.  The data were reduced according to standard procedures and calibrated
relative to 2MASS.  We very clearly show (Fig.~\ref{fig:ir}) that the single 2MASS
source is actually a blend of three sources: two relatively bright stars and one
fainter star.  The position of the pulsar is between one of the brighter stars and
the fainter star but inconsistent with all of them given the uncertainties (the
WHIRC astrometry is accurate to $\pm0\farcs1$).  Further investigation to subtract
the brighter stars and look for a source coincident with the pulsar is ongoing.

None of the five MSPs presented here are positionally coincident with $\gamma$-ray 
sources from the \textit{Fermi} Large Area Telescope (LAT) 4-Year Point Source
Catalog \citep[3FGL; see][]{Acero15}. This is not surprising given the relatively
small spin-down rates and much larger distances of these MSPs compared to the current
sample of $\gamma$-ray detected MSPs \citep{Abdo13}, as well as the strong diffuse
$\gamma$-ray background in the Galactic plane. The only exception is PSR J1957+2516,
which is relatively nearby ($D=3.1$ kpc) and has a moderately high spin-down rate
($\dot{E}=1.74\times10^{34}$ erg s$^{-1}$). To look for high-energy emission from this
MSP, we retrieved all Pass8 \textit{Fermi}
LAT data within 8$^{\circ}$ of the radio timing position of PSR J1957+2516 spanning
from the start of the mission through 2016 March 16. The data products were first
filtered with the {\it Fermi} Science Tools v10r0p5 using the event selection criteria
recommended by the {\it Fermi} Science Support Center\footnote{See
\url{http://fermi.gsfc.nasa.gov/ssc/data/analysis/scitools/} for details.}. Using
the counts, exposure, source maps, and live-time cube generated from these data
as well as a spatial/spectral source model taken from the 3FGL catalog, we carried
out a binned likelihood analysis to test for the presence of a $\gamma$-ray source
at the pulsar position. We further include a new source at the pulsar position modeled
with an exponentially cut-off power-law, as appropriate for a pulsar. The best-fit
model from the binned likelihood analysis results in a negative value for the test
significance (TS) for a source at the pulsar position. This is an indication that
the addition of a $\gamma$-ray source coincident with PSR J1957+2516 is not warranted
by the data, which in turn, suggests that this MSP does not produce $\gamma$-ray
emission that is detectable above the background level. {\bfref We also performed
a pulsation search for each of the 5 MSPs using the {\it Fermi} plugin for
{\tt tempo2}\footnote{\url{http://fermi.gsfc.nasa.gov/ssc/data/analysis/scitools/pulsar\_analysis\_appendix\_C.html}}.
For the pulsation search we selected events $\leqslant0.8^\circ$ from the MSP position and
used events with energies ranging from 0.1 GeV to 10 GeV, however there was no evidence of pulsations.}

\section{Conclusions}\label{sec:conc}
We have presented timing solutions for 5 MSPs discovered in the PALFA survey, which
continues to discover some of the most distant MSPs known. The most distant MSP presented
here is PSR J1938$+$2012, which has a DM-derived distance of 7.7 kpc. Four of the MSPs
were found to be in highly circularized binary orbits while the other system (PSR J1914$+$0659) is 
an isolated, partially-recycled pulsar indicating that it is likely the result of a
disrupted double neutron star system. One of the binary systems, PSR J1933$+$1726 has a
high-mass companion that is likely to be a CO white dwarf. Two others, PSRs J1906$+$0055
and J1938$+$2012, have low-mass companions that are likely to be He white dwarfs. The
remaining binary system (PSR J1957$+$2516) is likely to be a redback system. It is an
eclipsing system in a tight orbit with a low-mass companion and we have detected a
change in the orbital period. PSR J1957$+$2516 {\bfref was found to be close to a near-IR
source which appeared to be extended in archival data. Higher resolution images revealed
multiple point sources near the position of PSR J1957$+$2516, but further follow-up is ongoing to
determine of the pulsar is associated with one of the NIR point sources.} As is the case
for all PALFA-discovered MSPs, these MSPs show no signs of $\gamma$-ray emission significantly
above the background emission in the Galactic plane. This indicates that the PALFA survey
is complementary to the ongoing {\it Fermi} searches for MSPs.

\section*{Acknowledgements}
The Arecibo Observatory is operated by SRI International under a cooperative agreement 
with the National Science Foundation (AST-1100968), and in alliance with Ana G. M\'{e}ndez-Universidad
Metropolitana, and the Universities Space Research Association.
The pulsar group at UBC acknowledges funding from an NSERC Discovery Grant and
from the Canadian Institute for Advanced Research.
JvL acknowledges funding from the European Research Council under the European Union's Seventh
Framework Programme (FP/2007-2013) / ERC Grant Agreement number 617199.
JSD was supported by the NASA Fermi Guest Investigator program and by the Chief of Naval Research.
J.W.T.H. acknowledges funding from an NWO Vidi
fellowship and from the European Research Council under the European
Union's Seventh Framework Programme (FP/2007-2013) / ERC Starting
Grant agreement nr. 337062 (``DRAGNET'').
V.M.K. receives support from an NSERC Discovery Grant and Accelerator Supplement, 
from NSERC's Herzberg Award, from an R. Howard Webster Foundation Fellowship from the Canadian Institute for Advanced Study, the Canada Research Chairs Program, and the Lorne Trottier Chair in Astrophysics and Cosmology. {\bfref We thank an
anonymous referee for improving this manuscript.}

{\it Facilities:} \facility{Arecibo}, \facility{Jodrell Bank}

\bibliography{palfa_5msps}
\bibliographystyle{yahapj}

\end{document}